# Copper alloys deterioration due to anthropogenic action

A. Duran*, J.L. Pérez-Rodríguez*, L.K. Herrera*, M.C. Jiménez-de-Haro*, M.D. Robador*, A. Justo*, J.M. Blanes* and J.C. Pérez-Ferrer*

**Abstract**    Results are presented from several samples taken from leaves of the Pardon Portico of Mosque-Cathedral of Cordoba, where an alteration on their surface was detected. Metal samples analyzed using X-ray microanalysis and powder x-ray diffraction were predominantly constituted by copper with some amounts of zinc attributed to brass, whereas other samples were also constituted by copper, tin and lead attributed to bronze. Surface samples were analyzed using the same techniques. In addition Fourier transform infrared spectroscopy was also used. The main compound identified in all the surface of the leaves is copper chloride hydroxide (atacamite). Lead chlorides have also been found. These data show that the sudden alteration that appears may be attributed to the use of some cleaning product containing chloride. Other compounds detected in the surface were gypsum, quartz and oxalates coming from environmental contamination.

**Keywords**    Deterioration. Bronze. Brass. Cleaning. Human action.

## Deterioro de aleaciones de cobre por acción humana

**Resumen**    Se exponen los resultados de algunas muestras alteradas procedentes de la Puerta del Perdón, pertenecientes a la Mezquita de Córdoba. Algunas de las muestras de metal analizadas mediante difracción de rayos X y microanálisis por energía dispersiva de rayos X estaban constituidas por latón (cobre con pequeñas cantidades de zinc), mientras que otras estaban constituidas por bronce (cobre, estaño y plomo). La superficie de las muestras se analizó empleando las mismas técnicas. Asimismo, se usó espectroscopia de infrarrojos. El componente principal identificado en la superficie de la puerta es cloruro básico de cobre (atacamita). Además, se han encontrado cloruros de plomo. Estos datos demuestran que la repentina alteración de la superficie se puede atribuir al uso de productos de limpieza con alto contenido en cloro. Otros compuestos procedentes de la contaminación externa, como yeso, cuarzo y oxalatos también se detectaron en la superficie.

**Palabras clave**    Deterioro. Bronce. Latón. Limpieza. Acción humana.

## 1. INTRODUCTION

Most frequently used copper alloys are brasses (Cu-Zn), copper-nickels (Cu-Ni 70:30 or 90:10) and bronzes (Cu with Sn, Al or Si). Bronzes are used for the construction of different types of works of art, statues and monuments of cultural relevance, due to its colability and high stability to environmental corrosion[1].

The corrosion of bronze monuments has been studied by several workers[2-6]. This interest is mainly due to the increase awareness of air pollution damages to the cultural heritage. Some works have contributed to a better understanding on the reaction mechanisms of environmental deterioration[3].

The interaction with the atmosphere although slow and progressive leads to corrosion products that remain reactive along time. For instance, in environments polluted with sulphur compounds like urban atmospheres the first corrosion product produced is black cuprous sulphide than turns into blue cupric sulphide. Green basic copper salts can also be formed such as copper sulphates, basic copper carbonate or copper chloride in marine atmospheres. Natural patina may also contain small amounts of iron from silicon iron powder or soot[1, 7 y 8].

Degradation of the statues, as a result of corrosion, has resulted in parts of the surface becoming mainly black and others ranging through various shades of green[9] Different factors, such as accumulation of aerosols of different kinds, the increasing presence of traces of heavy metals with strong catalytic properties, residues from vehicles, and the increase of highly oxidizing substances such as non-metallic oxides and ozone[10], contribute to an unidentified decay mechanism of bronzes.

Frequently, the first reaction to occur is the oxidation of the base metal exposed to the atmosphere to produce cuprous ions. In a chloride polluted atmosphere, cuprous chloride is formed. This is an unstable compound that in the presence of oxygen and environmental humidity turns into basic copper chloride with simultaneous production of hydrochloric acid. The presence of this acid will induce a new cyclic reaction of copper until the total consumption of the base metal. Only if the cuprous chloride is eliminated these damaging cyclic process ("bronze disease") would be stopped[1, 7 y 8]

In metals belong to cultural heritage are easily altered by weathering effects. However, they may come from a previous treatment. The removal of the corrosion layer eliminates many of the soluble corrosion products, specially trapped ones that accelerate corrosion but it also removes the less soluble corrosion products that slow corrosion. Thus, the removal of the corrosion layer is a complex process and there is some discussion about the best method to clean bronze pieces of cultural heritage.

For the chemical cleaning of copper and copper alloys like bronze, the art work can be immersed in a cleaning solution or this latter can be also locally applied. The time of exposure to cleaning solution and its concentration would depend on the surface condition of the piece to be treated. There are certain unavoidable previous treatments such as degreasing of the surface and the removal of previous coatings. Once accomplished with this step, the piece of art must be taken off from the treatment vessel and rapidly washed. This step is relevant when acids solutions have been used as well as the further neutralization of acidic residues that must be eliminated by immersion in an aqueous solution of sodium carbonate. Finally, a wash must be performed. In the washing procedure, drinking water, distilled water or de-ionized can be indistinctly used to guarantee the complete removal of remaining traces of chemical cleaning reactants[1, 7 y 11].

When the treatment is carried out by a non-specialized company it may produce an undesired increase of the corrosion instead of the surface cleaning, especially when the product used contains chloride ions as it happened in the gateway studied in this work. Especially difficult is the removal of chloride ions from porous, pitted and corroded surfaces, because chloride ions usually concentrate superficially[12] . The results presented in this paper are an example of the risks that can be found on outdoor bronze, after surface cleaning by non - specialized personnel.

## 1.1. The Pardon Gateway of the Mosque-Cathedral of Cordoba

The Pardon Gateway is located in the North façade of Mosque-Cathedral of Cordoba (Fig. 1). The stone materials used in this monument have been previously studied by Barrios-Neira et al[13]. The gateway leaves were made with pine wood covered with bronze plates. The plates form a geometric-"Mudejar" (arabic style made in Christian time) drawing. The three following decoration are repeated on the leaves: a. *Of God is the empire of all*. b. *Deus* (the four letters are included in the free space between the cross and the perimeter of the shield). c) *a plant*. Also around leaves is written: "*saintly be the name of God*". More outstanding than the leaves is the knocker that is inspired in cordobes "Mudejar" style showing a marked influence of Arabic art from X century. Conversely, the inscription around it says: *"benedicto:s dominus deus: isra: el: quia: vi:"*.

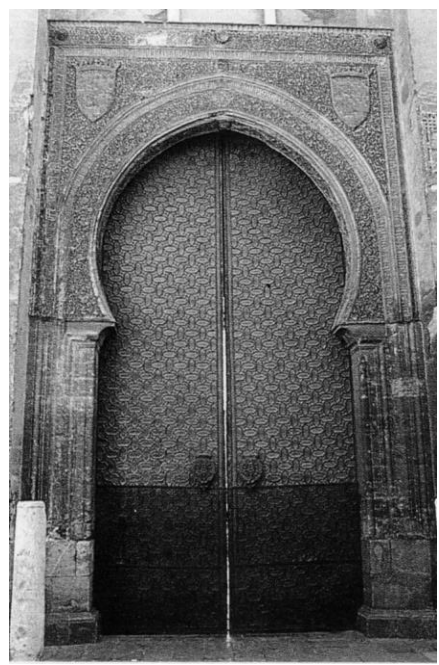

**Figure 1.** Pardon gateway of Mosque-Cathedral of Cordoba.

*Figura 1. Puerta del Perdón de la Mezquita-Catedral de Córdoba.*

The leaves that are dated on March 1377, are related with the "Almohade" Portico of the Seville Cathedral.

The gateway was rebuilt in 1739. The leaves were in a good conservation state until the middle of the XX century, when a sudden alteration began after being subjected to a cleaning process. The alloys and the rivet used to fix the plaques presented a high degree of oxidation. This deterioration was attributed to a wrong cleaning procedure with an inappropriate product, and it was necessary to stop the deterioration process as well as to eliminate the corrosion products.

## 2. EXPERIMENTAL

### 2.1. Materials

The following samples were taken:
— Sample 1 was taken from the reverse of left leaf, on sheets that surround the internal edge of the leaf.
— Sample 2 was taken from a separated fragment of a moulding.
— Sample 3 was taken from right leaf, located similarly to sample 1.
— Sample 4 is a powder taken from a caisson in the right leaf showing severe corrosion.
— Sample 5 is a powder taken from a rivet (Fig. 2).
— Sample 6 is a separated fragment (nail) from the top of the door. (Fig. 3).

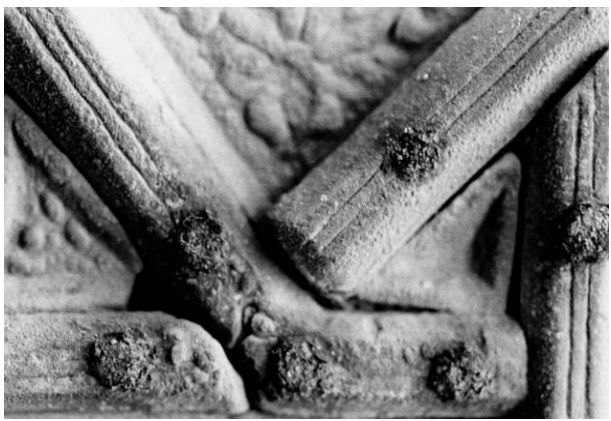

**Figure 2.** Details of the rivets.

*Figura 2. Detalles de los remaches.*

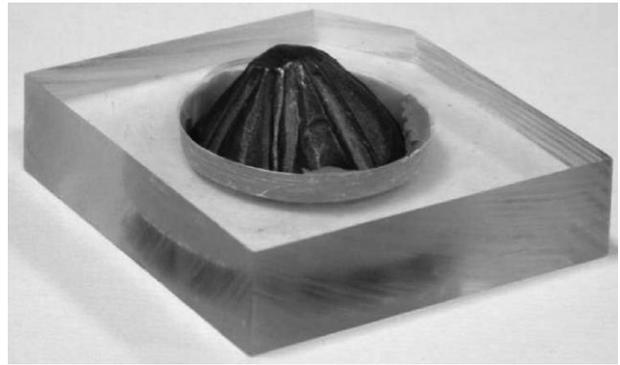

**Figure 3.** Separated fragment (nail) from the top of the door.

*Figura 3. Fragmento separado (clavo) de la parte superior de la puerta.*

### 2.1. Methods

The crystalline phases were characterized by X-ray diffraction (XRD) using a Siemens diffractometer model Kristalloflex D-5000. Göbel crystal has been used to study sample 6. The scanning electron microscopy (SEM) study was carried out with a Jeol JSMS5400 microscope equipped with an X-ray dispersive energy analyzer (EDAX). Fourier-transform infrared spectroscopy (FTIR) (Nicolet 510) was employed to determine the inorganic and organic compounds present in the encrustation.

## 3. RESULTS AND DISCUSSION

### 3.1. Copper alloys

In a first step, the alloys used in the leaves construction were characterized. The elemental analysis of sample 1 shows that the original material is constituted by copper and zinc (Fig. 4 a)). Punctual chemical analysis in this sample shows the presence of Zn (Fig.4 b)). The x-ray diffraction patterns of this sample support the presence of copper zinc alpha brass (figure not shown). The chemical composition and phases found for sample 3 are similar to those present in sample 1. These results confirm that these materials are constituted by copper zinc alpha brass. This alloy is present in a small part of the leaves, mainly in the edges.

The elemental analysis of sample 2 shows the presence of Cu, Sn and Pb (Fig.4 c)). A similar composition was found for sample 6 accompanied by Fe and small amounts of Zn. The X-ray diffraction patterns of this sample show that it is constituted by copper-tin alpha brass (Fig. 5). The x-ray diffraction patterns of this last sample has been carried out on the sample after cleaning by restaurators using a Göbel accessory that allow the direct study of irregularly shaped samples avoiding to obtain any powdered sample from the pieces, that it was not permitted by restaurators due to high artistic value of this piece. In figure 3 it is shown the montage used to obtain the x-ray diagram of these samples.

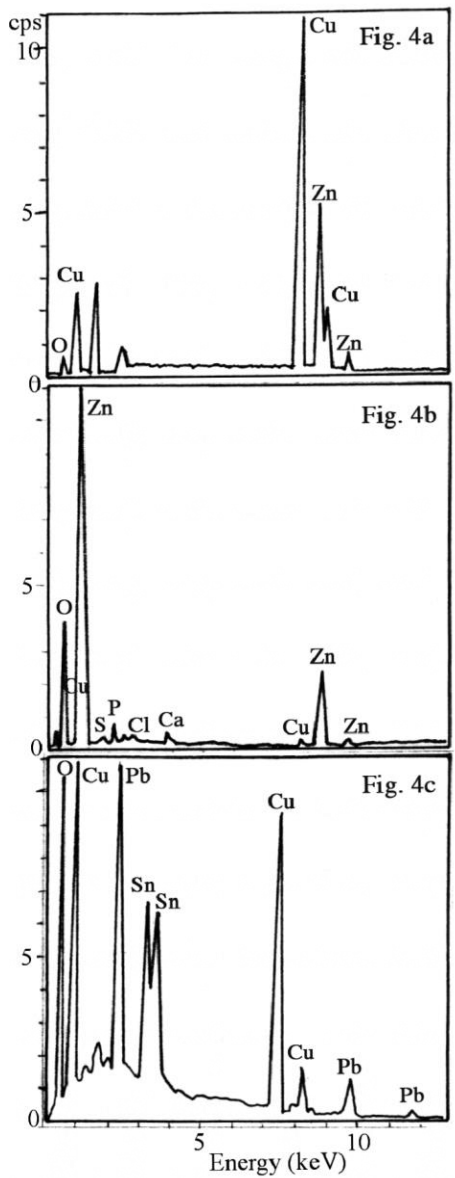

**Figure 4**. Elemental analysis of sample 1. (a) general and b) punctual); and sample 2 (c) general).

*Figura 4. Análisis elemental de la muestra 1 (a)general y b) puntual) y de la muestra 2 (c) general).*

The Göbel accessory consists in a multilayer crystal centred in the primary beam. The thickness of the layers increase along its length and a constant slope in the crystal is obtained. This form a crystal of parabolic geometry that may transform the divergent beam of X-ray in a parallel beam with high intensity: The parallel beam facilitate the study of very small pieces because the diffraction beams from the different crystals of the surface keep enough energy independently of surface morphology.

These results show that these samples are constituted by bronze as main component in the leaves of the gateway. The presence of lead is due to the manufacturing process frequently used (bronze cast) in the past.

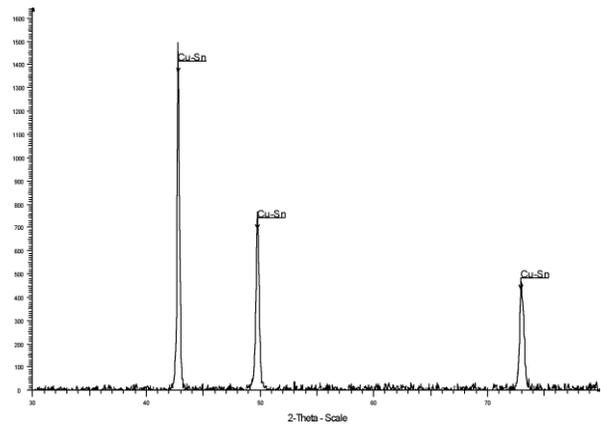

**Figure 5**. XRD patterns of unaltered material of sample 6 (Cu-Sn).

*Figura 5. Diagramas de difracción de rayos X del material inalterado de la muestra 6 (presencia de Cu-Sn).*

### 3.2. Altered materials

Elemental analysis of the samples 1 and 3 shows the presence of copper and oxygen. This last element is difficult to attribute to the oxidation of Cu because may form part different compounds present in the base material. However, the x-ray diffraction powder taken directly from the sample shows the presence of $Cu_2O$ (Fig. 6 a)).

Also, punctual elemental analysis in samples 1 and 3 shows the presence of Zn and O (Fig.4 b)). This finding may be attributed to formation of ZnO (zincite). When zinc is exposed out-doors initially form zinc oxides[14] that may be transformed into a basic zinc compound. It is possible that ZnO identified in this sample resulted from the corrosion of freshly exposed zinc. This phase has not been detected by X-ray diffraction.

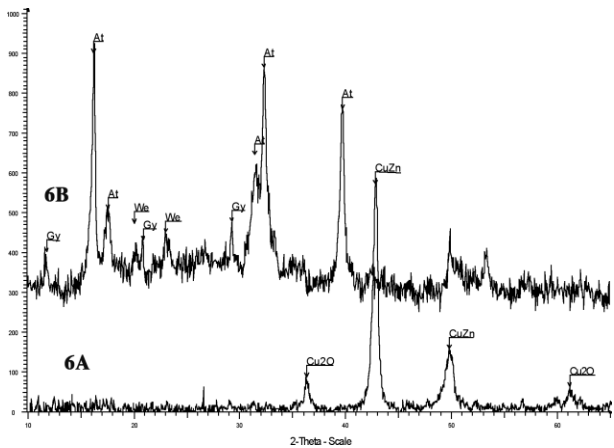

**Figure 6**. XRD patterns of oxidized material of samples 1 and 3 (Cu$_2$O and Cu-Zn) (**A**) and XRD patterns of corrosion and environmental contamination products of samples 1 and 3 (Gy=gypsum; At=atacamite; We=weddellite) (**B**).

*Figura 6. Diagrama de difracción de rayos X del material oxidado procedente de las muestras 1 y 3 (presencia de Cu$_2$O y Cu-Zn) (**A**) y Diagrama de difracción de rayos X de los productos de corrosión y de contaminación ambiental de las muestras 1 y 3 (presencia de Gy=yeso; At=atacamita; We=wedellita) (**B**).*

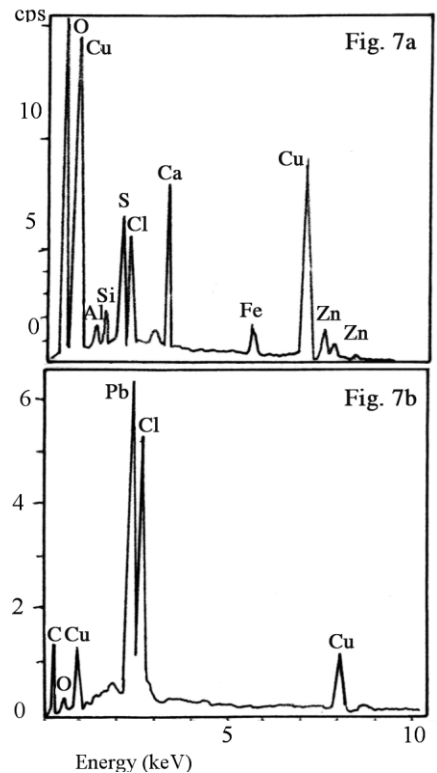

**Figure 7.** Elemental analysis of corrosion products a) from samples 1 and 3 (general); b) and sample 4 (punctual).

*Figura 7. Análisis elemental de las probetas de corrosión: a) de las muestras 1 y 3 (general); b) de la muestra 4 (puntual).*

Similarly results have been detected for tin (Sn and O), but romarchite (SnO) was no detected by X-ray diffraction due to low proportion. The presence of romarchite has been described in different works about alteration of tin[15].

All these chemical compounds are very well known and are frequently reported in the surface of all these alloys.

The elemental analysis of corrosion products from samples 1 and 3 show the presence of copper, chlorine, zinc, silicon, sulphur, calcium, iron, aluminium and oxygen (Fig. 7 a)) The XRD patterns shows that the sample is constituted by atacamite (basic copper chloride), gypsum and weddellite (Fig. 6 b)). Atacamite has been detected previously in copper and copper alloys exposed outdoors due to environmental contamination. This environmental contamination is also responsible of the presence of gypsum and quartz. Both may also come from the restoration process that was carried out around the gateway. The weddellite, also present, may be formed by reaction between environmental gypsum or calcite, with oxalate ions from dissolved acid. In urban areas oxalic acid is abundant in rain and mists[16] or it may be secreted by micro organisms, such as fungi and liquens[17].

The infrared spectra of different corrosion materials from these leaves show characteristic bands of atacamite, gypsum and quartz. It also appears a band at 1622 cm$^{-1}$ assigned to oxalate (Fig. 8).

The chemical analysis of the corrosion material from sample 2 shows similar composition that in samples 1 and 3, being also presents tin and lead although zinc is absent (figure not shown).

The sample 4 was taken from corrosion products that practically cover the two leaves of this gateway. The main component is the atacamite according with the results obtained by chemical analysis and XRD (figure not shown). Punctual analysis shows that some particles are constituted only by lead and chlorine (Fig.7 b)) confirming that the lead has been also attacked by chloride ions. This phase has not been confirmed by x-ray diffraction due to the scarce amount of the sample.

Atacamite, gypsum, quartz, weddellite and other deterioration products found in this work have been detected previously in other copper and copper alloys samples exposed to outdoors conditions. However, the main findings of this work are the high levels of chloride reported and the rapid deterioration observed after the cleaning process.

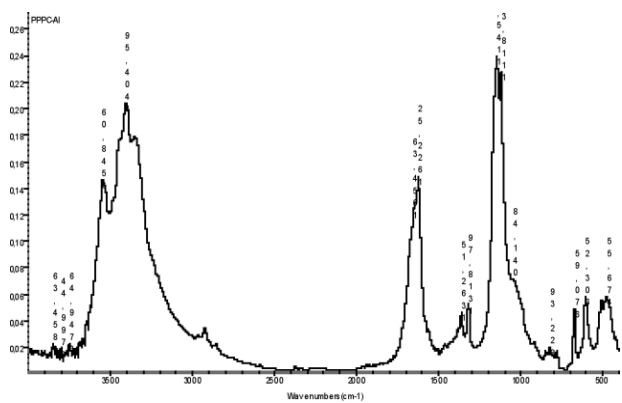

**Figure 8.** IR spectrum of corrosion (oxalate –band at 1622 cm$^{-1}$–; atacamite –bands at 3548 cm$^{-1}$, 3404 cm$^{-1}$, 822 cm$^{-1}$) and environmental contamination products (gypsum –bands at 1145 cm$^{-1}$, 1118 cm$^{-1}$, 670 cm$^{-1}$, 603 cm$^{-1}$–, quartz –band at 1041 cm$^{-1}$–).

*Figura 8. Espectro de IR de los productos de corrosión (oxalato –banda a 1622 cm$^{-1}$-; atacamita –bandas a 3548 cm$^{-1}$, 3404 cm$^{-1}$, 822 cm$^{-1}$-) y de los productos de contaminación medioambiental (yeso –bandas a 1145 cm$^{-1}$, 1118 cm$^{-1}$, 670 cm$^{-1}$, 603 cm$^{-1}$–; cuarzo – banda a 1041 cm$^{-1}$–).*

The chemical analysis carried out directly in the fragment of the sample 6 after cleaning still shows the presence of Cl, together with Cu, Sn, Pb, Fe and small amounts of Zn, showing the difficulty to remove ion chlorine.

## 4. CONCLUSIONS

The results presented here document the composition of the metallic substrate and the corrosion products on art work pieces from the gateway of Mosque-Cathedral of Cordoba. The corrosion products found in this study suggest not only the possible effect of environmental factors but also anthropogenic effects due to the use of inadequate cleaning procedures.

Although the presence of chloride in corrosion crusts on metals and alloys is frequently due to environmental exposure, in this particular case chloride ions are mainly provided by cleaning and leaching products from cleaning procedures accomplished by non-specialized technicians.

Two alloys have been found: one constituted predominantly by copper with some amounts of zinc that could be attributed to brass, and the other, constituted predominantly by copper accompanied by tin, lead and iron and rarely by a small proportion of zinc, that could be attributed to bronze.

The leaves were covered with a complex patina of corrosion products.

Copper corrosion has been enhanced by acid attack. The high concentration of chloride ions on the metal surfaces allows the formation of chloride compounds. The predominant corrosion products were green copper chloride hydroxides (mainly atacamite).

Other metal present in the alloys (as lead) also formed chlorides. Unusual deposits caused by environmental exposure such as gypsum, quartz and oxalates have been found in the external surface of the metal whereas chloride has been found in the corrosion products in the alloys composition.

The severe corrosion produced by chloride may be due to some chemical products containing this ion used for cleaning purposes.